\documentclass{aastex}

\slugcomment{To appear in AJ}

\shorttitle{A Solitaire Dwarf Galaxy}
\shortauthors{Pasquali et al.}

\begin{document}


\title{Discovery of a Solitaire Dwarf Galaxy in the
       APPLES Survey}


\author{Anna Pasquali}
\affil{Institute of Astronomy, ETH H\"onggerberg, HPF, CH-8093 Z\"urich,
       Switzerland \\
       ESO/ST-ECF, Karl-Schwarzschild-Strasse 2, D-85748 Garching
       bei M\"unchen, Germany}
\and
\author{S{\o}ren Larsen}
\affil{ESO/ST-ECF, Karl-Schwarzschild-Strasse 2, D-85748 Garching
       bei M\"unchen, Germany}
\and
\author{Ignacio Ferreras}
\affil{Institute of Astronomy, ETH H\"onggerberg, HPF, CH-8093 Z\"urich,
       Switzerland}
\and
\author{Oleg Y. Gnedin, Sangeeta Malhotra, James E. Rhoads, Norbert Pirzkal}
\affil{STScI, 3700 San Martin Drive, Baltimore, MD 21218, USA}
\and
\author{Jeremy R. Walsh}
\affil{ESO/ST-ECF, Karl-Schwarzschild-Strasse 2, D-85748 Garching
       bei M\"unchen, Germany}



\begin{abstract}
During the APPLES parallel campaign, the HST Advanced Camera for Surveys
has resolved a distant stellar system, which appears to be an isolated 
dwarf galaxy. It is characterized by a circularly symmetric 
distribution of stars with an integrated magnitude $m_{F775W}$ = 20.13
$\pm$ 0.02, a central surface brightness $\mu_{F775W} \simeq$ 21.33 $\pm$ 0.18 
mag~arcsec$^{-2}$ and a half-light radius of $\simeq$ 1.8 arcsec. The ACS and VLT 
spectra show no evidence of ionized gas  and appear dominated
by a 3 Gyr old stellar population. The OB spectral type derived for two
resolved stars in the grism data and the systemic radial
velocity (V$_{hel} \simeq$ 670 km~s$^{-1}$) measured from the VLT data give
a {\it fiducial} distance of $\simeq$ 9 $\pm$ 2 Mpc.
These findings, with the 
support of the spatial morphology, would classify the system among the 
dwarf spheroidal (dSph) galaxies. Following the IAU rules, we have named 
this newly discovered galaxy APPLES~1. An intriguing peculiarity of APPLES~1
is that the properties (age and metallicity) of the stellar content so far detected
are similar to those of dSph galaxies in the Local Group, where star formation
is thought to be driven by galaxy interactions and mergers. Yet, APPLES~1 
seems not to be associated with a major group or cluster
of galaxies. Therefore, APPLES~1 could be the first example of a field dSph
galaxy with self-sustained and regulated star formation and, therefore, would make
an interesting test case for studies of the formation and evolution of 
unperturbed dSph galaxies. 
\end{abstract} 




\section{Introduction} 
In the hierarchical framework dwarf galaxies are considered to be the building
blocks of larger baryonic structures, and observations indeed show that they are
the most abundant galaxies in the Universe. The SDSS
(York 2000) and 2dFGRS (Colless et al. 2001) surveys have shown
that the galaxy luminosity function keeps rising
down to the faintest magnitude probed (Blanton et al. 2001, Norberg et al. 2002),
with the faint end of the galaxy luminosity
function depending on environment. Hoyle et al. (2003) have found that
SDSS dwarf galaxies fainter than M$_r \simeq$ -15  are not generally
present in {\it voids} but preferentially populate {\it overdense regions} (i.e.
clusters; cf. Mo et al. 2004, Sabatini et al. 2003).
\par\noindent
Cold Dark Matter (CDM) theories
predict mass functions for subhaloes with a very steep slope at the
low mass end, virtually independent of the mass of the parent halo
(Kravtsov et al. 2004a). While this is true for the Virgo cluster
(Sandage et al. 1985, Moore et al. 1999), it is dramatically inconsistent 
with the observed galaxy content of 
the Local Group in which far fewer galaxies have been detected
than one would na\"ively expect for CDM cosmogonies (Klypin et al.
1999; Moore et al. 1999). A number of mechanisms for suppressing
dwarf galaxy formation have been included in CDM theories, such as
supernova-driven winds (Dekel \& Silk 1986) and reionization
(Bullock et al. 2000, Tully et al. 2002, Kravtsov et al. 2004b). 
These authors have suggested that reionization could shape the environment
dependence of the faint end of the galaxy luminosity function, so 
that galaxy clusters assembled before reionization would be dwarf-galaxy
rich in contrast with clusters formed after reionization (the ``squelching''
effect). 

Counts of dwarf galaxies in galaxy groups and clusters indicate that
they constitute the majority of the in-situ population, e.g. 80$\%$ in the
Virgo cluster (Binggeli et al. 1985) and 90$\%$ in the Local Group (Mateo 1998).
The best studied dwarf galaxies are in the Local Group, where they
have been classified (cf. Ferguson \& Binggeli 1994) into dwarf ellipticals 
(dE: with a compact
morphology and a high central surface brightness), spheroidals (dSph:
more diffuse structures with a low central surface brightness) and
irregulars (dIrr: with an irregular morphology and a significant content of
ionized and neutral gas). dEs and dSphs are ``gas-free'' systems
(Mateo 1998). The differences observed in the kinematic properties and chemical 
abundances (the latter being inferred from  the well-known
luminosity - metallicity relation for dwarf galaxies where
dIrrs are too metal-poor for their luminosity as compared
to dSphs) rule out that dIrrs naturally evolve into dSphs and dEs (Binggeli
1994, Grebel et al. 2003). Such a transition is rather induced by
the environment, via tidal
interaction with a closeby, more massive galaxy and via ram-pressure 
with the intergalactic medium in the Local Group, with the final effect
of depleting dIrrs of their original gas. Indeed, observations indicate
that dwarf ellipticals outnumber dwarf irregulars in the Local Group
(Mateo 1998), in Hickson compact groups (Zepf et al. 1997) and in the
Virgo cluster (Sandage et al. 1985), while dwarf irregulars in the 
field dominate the galaxy luminosity function at M$_B <$ -18 (Lilly
et al. 1996). 

Dwarf galaxies in the Local Group (as well as 
dwarf galaxies studied so far in the M81 group, Caldwell et al. 1998)
exhibit complicated star-formation histories and are characterized
by a mixture of old- and intermediate-age stars (dEs and dSphs)
at different metallicities, sometimes
with traces of a more recent starbust (few Myrs) like
in dIrrs (Grebel 1997, Mateo 1998, Tolstoy 2003, Grebel
et al. 2003). Although van
den Bergh (1994) suggested that dSph galaxies nearest to the
Milky Way are on average older than more distant ones, there is
no solid evidence for their star-formation history to be dependent on distance
from the Galaxy or M31 (Mateo 1998). Similarly, there is no clear
understanding of the mechanisms driving star formation in
the Local Group dwarfs. Galactic outflows
induced by supernova explosions, ram-pressure stripping, tidal
stripping, galaxy threshing, galaxy harassment and mergers are all thought
to play a role in controlling the star-formation activity
and gas content in satellite dwarf galaxies 
(Grebel 1997, Moore et al. 1999, Bekki et al. 2001). There
is no doubt that the dynamical evolution of a galaxy group ``washes
out'', in a sense, the pristine properties of the group dwarf galaxies.
Galaxy harassment in groups and clusters of galaxies can disturb
large disk galaxies and transform them into dwarfs (Moore et al. 1999,
Gnedin 2003a,b).
Galaxy interactions in compact groups are even able to produce an
ex-novo generation of dwarf galaxies, the so called tidal debris dwarf galaxies
(cf. Palma et al. 2002).

Have the truly pristine galaxy building blocks been found yet?
These are the unperturbed dwarf galaxies which most resemble
the galaxy building blocks at high redshifts.
In the nearby Universe, Tucana and Cetus, two dSph galaxies, are
located at about 880 kpc from the Milky Way and are considered
practically unbound to the Local Group. Unlike the satellite
dSphs, they do not show any recent burst of star formation and their stellar
populations date back to $\sim$ 10 Gyr ago (Mateo 1998, Grebel 1997).
Therefore, they may well represent primordial dwarf galaxies, which
have evolved as predicted by Dekel \& Silk (1986), i.e. dSphs descend from
gaseous systems which lost their gas because of a single burst of
star formation.

In this paper we report on the first detection of a gas-poor dwarf
galaxy which has undergone recent star formation and, yet, likely lies in the 
field, far away from any overdensity environment. 
We argue that this new galaxy, hereafter referred to as APPLES~1, 
may be located at a fiducial distance of $\sim$ 9 Mpc and with an
absolute magnitude of M$_{F775W} \simeq$ -9.6. This would make APPLES~1
possibly one of the faintest dwarf galaxies resolved in the field. 
Its case would then set important constrains
on feedback and on how star formation occurs in the smallest size scales of
dark matter. 

The discussion on the nature of APPLES~1 develops following the timeline along 
which more data became available for this galaxy: first, the survey 
images alone which allowed us to study the morphology and structure of APPLES~1,
and finally the ACS and VLT spectra which allowed us to estimate
the distance and stellar content of this galaxy.

\section{Observations}
\subsection{Data from the HST Advanced Camera for Survey}
Our team has led the ACS Pure Parallel Ly$\alpha$
Emission Survey programme (APPLES, PI Rhoads), a parallel programme approved
for Cycle 11. The scope of this programme was to collect pairs of direct
and grism images  of random fields at high galactic latitude in
order to study galaxy populations at redshifts greater than 1 (Pasquali 
et al. 2003b). The observational techniques and the data-reduction strategy
applied to imaging and spectroscopy are fully described in Pirzkal et al. (2004).
Briefly, the observations were carried out with  
the ACS Wide Field Channel (WFC), and consisted of pairs of direct and
grism images for a total exposure time of 1460 s in imaging and 3470 s in
spectroscopy. The direct images were acquired throught the F775W filter and 
were used to accurately measure the position of sources on the chips, which
fixes the zero point of the wavelength calibration of their grism spectra
(cf. Pasquali et al. 2003a). The direct images were reduced using CALACS (Hack 1999)
and PyDrizzle (Sparks et al. 2001). The ACS pipeline was only partially run
on the grism images, to allow the correction for bias, dark current and for
the different detectors gain. Spectra were extracted from each grism
image and calibrated
with the ST-ECF package aXe (Pirzkal et al. 2001), which has built-in
the correction for flat-field, the wavelength and flux calibrations.
Multiple spectra of the same source were then medianed.
\par
In the direct images, we detected an isolated and compact stellar system.
The field appears sparsely populated by galaxies, with very few 
field stars; the stellar system is characterized by a circularly 
symmetric concentration of numerous faint, point-like sources. 
In Figure 1, we compare the image of this stellar system
taken with the ACS/WFC (9$'' \times$ 9$''$, middle panel) with the DSS image of the
Galactic GC Pal 13 (5$' \times$ 5$'$, left-hand side panel) and the V-band image of the dwarf 
galaxy And V (3$' \times$ 3$'$, right-hand side panel) published by Armandroff et al. 
(1998). The stellar system is itself
visible on the DSS as a faint and unresolved point source. Following the IAU
rules, we have named this newly discovered system APPLES~1. 

\subsection{Data from the VLT FORS2}
We have observed APPLES~1 with FORS2 on the ESO VLT/UT4 as part of the Director General
Discretionary Time (Program ID 271.D-5043A, PI: Pasquali), on August 30 2003. 
Specifically, we have
performed low-resolution spectroscopy in the range 3500 \AA\ - 9000 \AA\ 
through the GRIS$_-$300V grism and for a total exposure time of 1.5 hours. The adopted
slit was 1.6 arcsec wide resulting in an effective dispersion of 3.3 \AA/pix. The data have
been reduced with the IRAF packages TWODSPEC and ONEDSPEC, achieving an
accuracy of 0.3 \AA/pix in the wavelength calibration, and 7$\%$ in the flux calibration.
We present and discuss the ACS/grism and FORS2 spectra in Sect. 4.

\section{Photometric properties} 
Figure 2 shows the location of APPLES~1 ({\it l} = 348$^o$.5 and {\it b} = -65$^o$.1)
in comparison with the Galactic OB associations, open and globular clusters 
(top panel; Melnik et al. 1995, Lynga 1987, Harris 1996)
and the Local Group members (bottom panel; Mateo 1998) in galactic coordinates. 
Galactic OB associations, open
and globular clusters are represented as crosses, open triangles and circles, while the Local Group
galaxies are shown as open triangles. The Sculptor dwarf galaxy and two other galaxies
in the Sculptor Group (NGC~55 and NGC~7793) are represented
with asterisks and APPLES~1 with a filled circle. The dashed and solid lines trace the
Sagittarius and Magellanic streams respectively (Martinez-Delgado et al. 2003, Kunkel 1979).
APPLES~1 appears to be far away from any Local Group member
and from the Sagittarius stream, less so for the Magellanic stream; the closest
galaxies ($\sim$ 13$^\circ$ away from APPLES~1) are NGC~55 and NGC~7793 at a distance of 1.9 and 
4.3 Mpc respectively. Assuming that APPLES~1 were at same distance in either cases,  
its projected distance from NGC~55 and NGC~7793 would be $\simeq$ 450 and 980 kpc, respectively. 
\par

Photometry of point sources in APPLES 1 was carried out with the
DAOPHOT package (Stetson 1987), running within IRAF. Point
sources were first detected by running the DAOFIND task with
a detection threshold of 3 sigma above the background noise.
Aperture and PSF photometry were then obtained with the PHOT
and ALLSTAR tasks. Following the first pass of ALLSTAR photometry,
the DAOFIND task was run a second time on the residual image
produced by ALLSTAR in order to detect additional objects which
might have been missed in the first pass due to crowding,
especially near the centre of APPLES 1. The two object lists
were then merged and the PHOT $+$ ALLSTAR sequence repeated.
Finally, the photometry was calibrated to the VEGAMAG system
by measuring a few isolated stars in the image in a $0''.5$
aperture and applying the zero-points in De Marchi et al. (2004). 
The observed luminosity function of the detected sources is shown
in Figure 3.

The surface brightness profile of APPLES 1 was measured by
obtaining aperture photometry in concentric apertures, centered
on the object. The background was measured in an annulus with
an inner radius of 100 pixels (5 arcsec) and 50 pixels wide.
The number counts within each annulus was normalized to the
area of the annulus and then converted to VEGAMAG surface
brightness using the same photometric zero-points as for the
PSF photometry. The total integrated magnitude within a
5\arcsec\ radius is M$_{\rm VEGA}$(F775W) = $20.13\pm0.02$.  The surface 
brightness profile is shown in Figure 4, 
where the ``bump'' at
$0''.3$ is due to the presence of resolved sources within
APPLES 1.  The surface brightness 
has dropped to half its central value at a radius of $\sim0''.4$, 
and half of the total light is contained within an aperture with a radius
of $1''.78$.
\par\noindent
We have fitted to the observed surface brightness profile a Sersic profile
(solid line) of the form (Sersic 1968):
\par\noindent
$\Sigma$ = $\Sigma_o e^{-(r/r_s)^{(1/n)}}$
\par\noindent
The best fit is obtained for a scale length $r_s$ = 0$''.$35 $\pm$ 0$''.$14,
a central surface brightness $\Sigma_o$ = 21.33 $\pm$ 0.18 mag~arcsec$^{-2}$ 
and a Sersic index $n$ = 2.30 $\pm$ 0.29.

\subsection{First guess on the nature of APPLES~1}
In the hypothesis that APPLES~1 were a Galactic open cluster since a small number
of these clusters are located at $|b| \geq$ 50$^\circ$ (cf. Lynga 1987), it would turn out 
to be improbably small and poor. The farthest known Galactic open cluster is 
at 8 kpc from the Sun: at this distance, APPLES~1 would be 0.08 pc in size and
have less than a solar luminosity.
\par\noindent
If we assume that APPLES~1 is, instead, a 10 Gyr old star cluster like a Galactic
globular cluster (GC),
then its brightest stars are expected to be red giants with $M_I \simeq$ -3. Since the
brighter sources in APPLES~1 are detected at $m_{F775W}$ = 24 mag, the distance
modulus of APPLES~1 would turn out to be (m - M) $\sim$ 27, corresponding to a distance of
$\simeq$ 2.5 Mpc. The limited statistics on the number of detected RGB-tip stars 
also allows for a smaller distance; a value of 1.5 Mpc would
be still consistent with the magnitude distribution of the observed stars. At this distance:
\par\noindent
{\it i)} The integrated M$_{F775W}$ of APPLES~1  would be $\sim$ -7, 
or M$_V$ $\sim$ -6 [assuming (V - I) = 1., cf. Renzini \& Fusi Pecci (1988)], well 
within the range of luminosities spanned by Galactic GCs though about 1 mag fainter 
than the mean.
\par\noindent
{\it ii)} The half-light radius $R_h \simeq$  1$''$.8 would correspond to
22 pc, a typical value among the Galactic GCs in the outer halo (e.g. Pal 5: 19 pc,
Pal 14: 24 pc and NGC~2419: 17 pc, Harris 1996). 
\par\noindent
{\it iii)} APPLES~1 would then classify as an {\it Intergalactic Globular Cluster},
belonging either to the field or to the Sculptor Group. If APPLES~1 were in the field,
it would likely represent an example of a pristine intergalactic globular
cluster formed out of primordial density fluctuations as predicted by  
Peebles \& Dicke (1968).  
If, instead, APPLES~1 were member of the Sculptor Group, it would definitively
confirm the existence of globular clusters in the intergalactic medium of galaxy
clusters, stripped from their parent galaxy by galaxy interactions 
as shown by Muzzio et al. (1984) and Harris \& Pudritz (1994). Hydrodynamical
simulations of globular cluster formation predict that massive star clusters are
born near the centers of dwarf galaxies at redshift $z \sim$ 3 - 5 and may escape
if their parent galaxies are disrupted by massive neighbors (Kravtsov \& Gnedin
2003).
\par
In the hypothesis that APPLES~1 were a globular cluster, would it survive dynamical
evolution over a Hubble time? Figure 5 shows
the location of APPLES~1 in the plane $L_V - R_h$ along with
the Galactic globular clusters and the dwarf galaxies in the Local
Group. The arrow in the plot indicates how APPLES~1 would move through the
diagram by increasing its distance. 
The structural parameters measured for APPLES~1 in the ACS image at an
assumed distance between 1 and 2.5 Mpc place APPLES~1 among the Pal GCs beloging
to the Galactic outer halo. At this distance, the relaxation time (cf. Spitzer 1987) 
of APPLES~1 turns out to be $\sim$ 8 Gyr, too large to affect its structure.
Only if its distance were smaller than 300 kpc, APPLES~1
would lie in the zone of destruction (the shaded area in Figure 5), where the
two-body relaxation would disrupt APPLES~1 in a Hubble time.
\par\noindent
For a total luminosity L$_V$ = 2.2 $\times$ 10$^4$ L$_{\odot}$ (assuming M$_V$ = -6 as
in point {\it i)} and a half-light radius of 22 pc, the
average luminosity density in APPLES~1 is L$_V/$(4$\pi$$R_h^3$/3) $\simeq$ 
0.38 L$_{\odot}$~pc$^{-3}$.
This value matches the lower luminosity
density derived for galactic GCs (i.e. the Pal GCs in the outer Galactic halo, Harris 1996) 
as well as the higher value estimated for dwarf galaxies in the Local Group (Mateo 1998).
Indeed, larger distances, between $\sim$ 10 Mpc and $\sim$ 100 Mpc, would move APPLES~1 
in the $L_V - R_h$ plane to among the dwarf galaxies. 
\par
At this point, it is clear that photometry alone can not give a straight answer to the
question of the nature of APPLES~1. It definitively rules out the possibility for
a galactic open cluster, but it does not discriminate between the scenario of a globular 
cluster and that of a dwarf galaxy. The final
answer can come only from spectroscopy, which allows to measure the radial velocity
of APPLES~1 and to estimate the age and metallicity of its stellar population.

\section{Spectroscopic confirmation of APPLES~1}
Optical grism spectra have been extracted for the brightest sources in APPLES~1
with the ST-ECF package aXe (Pirzkal et al. 2001). These sources  are numbered in
Figure 6 and their spectra are shown in Figure 7. 
While Sources $\#$ 2 and 3 have a stellar PSF (FWHM $\simeq$ 1.6 pixels), Source $\#$ 1
appears to be more extended, either because of superposition with nearby
stars or because it is a non-resolved star cluster. 
Because of the low resolution of the G800L
grism (40 \AA/pix, cf. Pasquali et al. 2003a), no spectral features are resolved 
and only the slope of the continuum can be used to estimate the spectral type. 
We have measured the continuum slope
between 6000 \AA\/  and 9000 \AA\/ and fitted it against the template spectra in the
atlas of Pickles (1985). The results are shown in Figure 7, where
the solid lines are the best-fitting templates. Source $\#$ 1 is well fitted by the
spectral types F5V, F5IV, F5III and F8I; 
the dashed and dotted lines
represent the A5 and G5 spectral types, respectively and are shown for comparison. 
The best-fitting templates for Source $\#$ 2
are the O5V, O8III and O9V spectral types, although also the B5V and B9III spectra
(dashed and dotted lines) seem to match the observed data reasonably well. Source 
$\#$ 3 is reproduced 
by the spectral templates B3V, B3III, B5III, B6IV (solid lines) and O5V (dashed line), 
while the A5V spectral type seems to underestimate the continuum for wavelengths bluer 
than 7000 \AA. We can now combine the photometry and the spectral types of these
Sources to estimate the distance of APPLES~1. In particular, we focus on Sources $\#$ 2 
and 3 which are point-like. Their F775W magnitudes are 24.08 $\pm$ 0.05 and
24.21 $\pm$ 0.05 respectively; given an average $(V-I)$ color of -0.5 mag for
OB stars (cf. Zombeck 1992), we infer a magnitude in V band of $\simeq$ 23.58 and
23.71 for Source $\#$ 2 and 3 respectively. O stars span a range of luminosity class
from $M_V$ = -5.4 (giants), -5.7 (main sequence) to -6.4 (supergiants), which implies
for Source $\#$ 2 a distance modulus $(m-M)_V$ between $\simeq$ 29.0 mag and 30 mag.
The luminosity class of B stars varies between $M_V$ = -1.1 (main sequence) to
$M_V$ = -6.9 (supergiants), hence the distance modulus of Source $\#$ 3 lies between
$\simeq$ 24.8 mag and 30.6 mag. This rules out that Sources $\#$ 2 and 3 are foreground
Galactic OB stars; on the other hand, both field OB stars and OB associations/open clusters
have not been detected at the Galactic latitude of APPLES~1 (cf. Ma\'iz-Apell\'aniz et 
al. 2003, also Figure2). It is then very likely that Sources $\#$ 2 and 3 have the
same distance modulus of $(m-M)_V$ = 29.6 $\pm$ 0.5 mag and lie at a distance of
8.5 $\pm$ 2 Mpc, the uncertainty here due to their luminosity class which can not be
determined from their grism spectra. The same comparison between the observed 
magnitude and spectral type of Source $\#$ 1 and the tabulated average luminosities
of F stars indicates that Source $\#$ 1 is an extragalatic source at the distance
of Sources $\#$ 2 and 3.
\par
The FORS2 spectrum of APPLES~1 is plotted in Figure 8. Since the spectrum was acquired
with a seeing of $\simeq$ 0$''$.5 and with a slit 1$''$.6 wide, it originates in the very
core of APPLES~1 which includes Sources $\#$ 1, 2 and 3. This spectrum   
is characterized by quite strong Balmer 
lines in absorption of which H$\epsilon$ is blended with the Ca H line. The Balmer lines
and the overall slope of the continuum can be nicely fitted with a combination of Pickles'
F2V and F5V spectral types. The most significant discrepancy is in
the Ca K line which is much less pronounced in the spectrum of APPLES~1 (cf. Figure 9). 
Since we have used
Pickles' templates with solar metallicity, the discrepancy in Ca K indicates a sub-solar
metallicity and/or the presence of OB stars (as already pointed out by the ACS grism spectra)
whose continuum would fill the Ca lines in the F spectral types.
\par\noindent
From the Balmer lines in the FORS2 spectrum we have measured a
radial velocity V$_{hel}$ = 674 $\pm$ 30 km~s$^{-1}$. If this were to be entirely due 
to the Hubble flow, then the distance of APPLES~1 would be 9.6 $\pm$ 0.5 Mpc assuming 
H$_o$ = 70 km~s$^{-1}$~Mpc$^{-1}$, nicely consistent with the spectroscopic distance
derived above. Deviations from the Hubble flow, such as random
peculiar motions as large as $\sigma_v \simeq$ 40 km~s$^{-1}$ and peculiar velocities
in the local field (amounting to $\simeq$ 20 km~s$^{-1}$ at the Galactic latitude
of APPLES~1) are known to occur within 5~Mpc from the Local Group (cf. Karachentsev et
al. 2003). As shown earlier, the spectral types of the resolved sources in APPLES~1
place this galaxy at a distance larger than 6.5~Mpc. However, if the deviations 
from the Hubble flow measured up to 5~Mpc were still applicable, they would increase
the uncertainty in the kinematic distance of APPLES~1 to $\simeq$ 1 Mpc. The
spectroscopic and kinematic estimates overlap for a {\it fiducial} distance of APPLES~1
of $\simeq$ 9 $\pm$ 2 Mpc. The uncertainty here takes into account the inability
of deriving a precise luminosity class for Sources $\#$ 2 and 3 from the ACS grism spectra 
and the deviations from the Hubble flow. At this distance, 
the absolute F775W VEGAMAG magnitude of APPLES~1 would 
be $\simeq -$9.6 (with a distance modulus of $\simeq$ 29.8 mag). Assuming
a (V $-$ I) colour of 0.6 mag as for a main-sequence F5 star (Zombeck 1992),
APPLES~1 would have M$_V \simeq -$9 mag.

\subsection{Lick indices}

More quantitative estimates of the luminosity-weighted
age of the stellar population in the core of APPLES~1 can be made by comparing
Lick indices with model predictions. As shown by Worthey (1994) and Trager et
al. (2000), Lick indices
have to be determined with high accuracy to set firm constrains on
the age and metallicity of stellar populations. This requires, in turn,
spectroscopic data with high S/N, typically higher than 30 - 40.
The S/N ratio in the line for the Balmer lines in the spectrum of the core of APPLES~1
is $\sim$ 35, and thus allows us to derive with some accuracy  
the Lick indices in  H$\beta$, H$\gamma_A$ and H$\delta_A$ 
as defined  by Worthey et al. (1994) and Worthey \& Ottaviani (1997). 
Figure 10 compares these H$\beta$, H$\gamma_A$ and
H$\delta_A$ indices with the theoretical values predicted by Bruzual \& Charlot (2003)
models as a function of age and metallicity. The solid
line indicates models with solar metallicity, whereas the dotted and dashed
lines represent models with LMC and half LMC metallicity. The horizontal
solid line corresponds to the measured Lick index and the shaded areas
visualize the 1 $\sigma$ confidence levels. While it
should be emphasized that the Lick system is not optimized for analysis of
young stellar populations (Worthey et al. 1994), all of the plots
indicate a luminosity-weighted mean age for the core of APPLES~1 on the order of 1 - 2 
Gyr. The higher-order Balmer lines (H$\gamma$, H$\delta$) indicate
somewhat younger ages (a few 100 Myr) than H$\beta$,
possibly (at least in part) due to a mix of stellar ages in the core of APPLES~1.
At shorter wavelengths, the younger population would tend to dominate,
resulting in younger luminosity-weighted age estimates from higher-order
Balmer lines (e.g. Schiavon et al. 2004).
\par\noindent
Although the S/N ratio in the Mg absorption feature at around 5180\AA\ 
is significantly lower than 
in the Balmer lines, we have also attempted to measure the Mgb index (cf.
Worthey et al. 1994 and Worthey \& Ottaviani 1997) and 
plotted it in Figure 10 against the predictions of Bruzual \& Charlot
(2003) for different ages and metallicities. Similarly to the H$\beta$ index, 
the Mgb index is consistent with an age
of 1 - 5 Gyr. Unfortunately, the Fe5270 and Fe5335
lines are characterized by the same S/N value as measured for the continuum,
preventing us from any reasonable determination of the $<$Fe$>$ index. We then
qualitatively conclude that the core of APPLES~1 is metal poor possibly with  
super-solar $\alpha$-element abundance ratios. 

\par
The age derived from the Lick indices and the radial velocity measured from the
Balmer lines clearly rule out the early hypothesis that APPLES~1 is an intergalactic
globular cluster (cf. Sect. 3.1). At the fiducial distance of about 9 Mpc,
the half-light radius of APPLES~1 would be $\simeq$ 78 pc, a factor of about 3 larger
than the half-light radius of Galactic GCs belonging to the Galactic halo, and comparable 
to the core radius of the dwarf galaxies EGB~0427$+$63 and DDO~210 in the Local Group. 
Therefore, we classify APPLES~1 as a dwarf galaxy, and the absence of gas emission
lines (e.g. H$\beta$, [OIII] $\lambda$5007 and H$\alpha$) in the ACS grism and
VLT/FORS2 spectra points to a spheroidal system. With M$_V \simeq$ -9.4 mag
APPLES~1 would be among the faintest Local Group dwarf galaxies such as Draco and Ursa Minor, 
and the faintest dwarf galaxies observed in Virgo (Sabatini et al. 2003). With
$R_h \simeq$ 78 pc which is even smaller than the core radius typical of
dwarf galaxies in the Local Group, APPLES~1
would also turn out to be the smallest dwarf galaxy ever resolved.

\subsection{Chemical enrichment}

So far we have compared the spectroscopic data from the core of APPLES~1 with 
either stellar data or simple stellar populations which correspond
to a single age and metallicity.
Except for small systems such as globular clusters, formed over
timescales much shorter than those expected for the enrichment of the
ISM, one should expect the stellar populations to be composed of a 
range of ages and metallicities. However, the age-metallicity 
degeneracy (Worthey 1994) poses a major hurdle in the estimate of 
the star formation history from integrated photo-spectroscopic 
data. A model-dependent approach 
involves the generation of a composite
stellar population according to a star formation history which 
consistently determines the age and metallicity distributions of the stellar 
component. We have followed the galactic chemical enrichment model
described in detail in Ferreras \& Silk (2000; 2003) and 
references therein. 
This model reduces the process of star formation to three 
mechanisms, namely:
\begin{itemize}
\item Infall of primordial gas following a generic function. We assume
  this function to be a delayed exponential: 
  $f(t)\propto\Delta t(z_F)\exp(-\Delta t(z_F)/\tau_f)$,
  where $\Delta t = t - t(z_F)$. Hence, 
  the infall rate is characterized by two parameters, the 
  formation redshift ($z_F$) which fixes the epoch when infall
  starts, and the timescale $\tau_f$. 
  This function rises quickly to its maximum value at $\Delta t=\tau_f$ and
  then declines on a slower timescale. 
  After a time $\Delta t_{90}=3.9\tau_f$, 90\% of the total gas content 
  has fallen on the galaxy 
\item The infall of gas feeds the gas reservoir which processes gas
  into stars via a Schmidt law: 
  $\psi (t)=\nu\rho_T\Big(\rho_g/\rho_T\Big)^{1.5}$, where
  $\nu$ is the star formation efficiency and $\rho_T$ refers to the
  total baryon content. $\rho_T$ is just a normalization term included
  so that $\nu$ is given in Gyr$^{-1}$. The inverse of the star
  formation efficiency is the star formation timescale.  
  Values of $\nu$ are usually low for
  late-type systems ($\nu\sim 0.1-1$~Gyr$^{-1}$, e.g. Boissier et al.
  2001)  and higher for elliptical galaxies ($\nu\geq 5$~Gyr$^{-1}$,
  Ferreras \& Silk 2000).
\item Outflows play an important role in the chemical enrichment 
  history as they ``modulate'' the yield and alter the gas content
  in the reservoir. We include a free parameter ($0\leq\beta\leq 1$)
  which describes the fraction of gas and metals ejected in outflows
  from evolved stars. These outflows are mainly dominated by
  type~II supernovae ejecta.
\end{itemize}
Type Ia SNe are included following Ferreras \& Silk (2002) who assume
the prescription of Greggio \& Renzini (1983) for a single
degenerate progenitor.

Therefore, for a given choice of the four parameters described above, namely:
$(\tau_f, z_F, \nu, \beta)$, we solve for the star formation history (SFH) 
which gives us the age and metallicity distribution of the resulting 
composite stellar populations. This SFH is convolved with the latest
Bruzual \& Charlot (2003) population synthesis models to generate
a spectral energy distribution (SED) with 3\AA\  FWHM resolution in the region 
of interest. A comparison of the model SED and the VLT spectrum of 
the core of APPLES~1
is then performed via a $\chi^2$ test in which the model SED is smoothed
to the lower resolution of the observations (FWHM$\sim 7$\AA )
and normalized to the same flux integrated in the spectral region of 
interest. 
The observed SED is in the range $3110<\lambda/\AA\ <9800$, sampled at
$3.3$~\AA\  per pixel, 
but we clipped it to a shorter interval to reduce noise and to improve the
accuracy of our age and metallicity estimates. Furthermore, we resampled
the spectrum, doing a 3~pixel rebinning in order to work with a pixel 
size roughly larger than the FWHM.
We mainly truncated the
spectrum at the red end, using a spectral range $3700<\lambda/$\AA\ $<6000$.
In the analysis of the likelihood function, we assumed a 7\% uncertainty per 
(unbinned) pixel in the spectrum. 
We chose a large range of star formation histories solving the
chemical enrichment model described above for a wide range of parameters.
We explored 11 different values of the outflow fraction from 
$\beta=0$ to $1$ in steps of $\Delta\beta=0.1$. For each choice, we
computed a grid of $32\times 32\times 32$ models for a wide range of 
star formation efficiencies:
$-2\leq\log(\nu/{\rm Gyr}^{-1})\leq 0$, formation times:
$2\leq t(z_F)/{\rm Gyr}\leq 13$ and infall timescales:
$1\leq\tau_f/{\rm Gyr}\leq 10$. 

Table~1 shows the best fit for the parameters
used to describe the star formation history as well as more
physical parameters such as the average and RMS of the stellar
ages and metallicities. The error bars are given at the 
$3\sigma$ confidence level, but we warn the reader to consider
these results as rough estimates. Indeed, one should notice that
the small uncertainty in the average age or metallicity is not
very meaningful since the spread in ages and metallicities is
rather large. Nevertheless, the strength of the prediction lies
in the need for a wide distribution of ages and metallicities.
The best model requires a rather low
star formation efficiency, comparable to those found in low-mass
disk galaxies. The combination of a low efficiency with an
extended infall timescale results in a significant ongoing 
star formation, suggesting that the core of APPLES~1 is still in its starting
stages of formation. Strong constraints cannot be imposed on
some of the parameters such as the outflow fraction or the infall
timescale. The star formation efficiency and especially the 
formation redshift -- which roughly determines the average 
age of the stellar populations -- are significantly constrained.
Figure~11 gives the age and metallicity histograms (left panels)
and their cumulative distribution (right panels). The solid and dashed
lines correspond to two extreme star formation histories within 
the $3\sigma$ confidence levels, labelled as H1 and H2. In both scenarios
we assume $\tau_f=5$~Gyr. H1 (solid line) represents a late ($z_F=1.5$) 
and low-efficiency formation process ($\nu =0.3$~Gyr$^{-1}$) with no
outflows, whereas H2 (dashed line)
corresponds to an early ($z_F=3$) and highly efficient formation 
history ($\nu =3$~Gyr$^{-1}$) with some outflows ($\beta =0.2$).
Both models yield a significant fraction of stars younger
than $3$~Gyr, with an average age around $2-4$~Gyr. 
This average age corresponds to a Main Sequence turnoff at around
an F stellar type, in agreement with the previous comparison of
APPLES~1 with stellar spectra.
The age distribution allows
for a non-negligible ongoing star formation, as seen by the presence
of OB stars. The star and its ($3\sigma$) error bar in the left panels 
represent the best fit for a 
comparison of the APPLES~1 spectrum with a grid of simple stellar
populations (i.e. single age and metallicity), drawn from the same
synthesis models of Bruzual \& Charlot (2003).
The discrepancy between a composite model and a simple stellar population
is remarkable. This should serve as a cautionary
warning in the use of simple stellar populations to explore the
star formation histories of unresolved stellar populations.
From a statistical point of view the analysis with a simple 
stellar population gives an equally acceptable fit, with a 
reduced $\chi^2\sim 1$. 
The strong Balmer absorption lines seen in the SED of APPLES~1
are characteristic of ~1 Gyr old stars, as seen in Figure~10. 
Hence, the use of simple stellar populations imposes a single age, 
which creates the significant shift towards younger ages in order 
to recover these strong absorption lines. The weakness of the 
observed Ca $K$ line favors models with subsolar metallicity,
as opposed to the SSP prediction.
Furthermore, the presence of OB stars along with the FORS2 integrated
spectrum matching a cooler F spectral type (see \S4) suggests a significant
range of ages. All this evidence points towards an extended 
period of star formation as implied by the chemical enrichment models.

Figure~12 shows the best fit SED (dashed line) along with the observed
VLT/FORS2 spectrum (solid line). The residuals are given in the bottom
panel, in the same units, namely
$10^{-18}$~erg~s$^{-1}$~cm$^{-2}$~\AA$^{-1}$.   The  inset  shows  the
range  of star  formation  histories within  the $3\sigma$  confidence
level (i.e. histories H1 and H2 defined above).  
The dashed  lines give  the  range of  star formation  rates,
whereas the solid lines trace  the evolution of the metallicity of the
ISM.  The point  gives the  marginalized average  age  and metallicity
along with the RMS shown as error bars.  The best fit of the models to
the  spectrum of  APPLES~1 suggests  ongoing star  formation  which --
scaled to  an absolute luminosity $M_V\sim  -9$ -- should  be of order
$\psi\sim  10^{-3} M_\odot$~yr$^{-1}$. However,  the lack  of emission
lines  would rather  suggest the  contrary.  A  rough estimate  of the
H$\alpha$ and  [OII] luminosities expected  from star formation  (e.g. 
Kennicutt 1998)  readily shows that  the VLT spectrum of  the APPLES~1
core strongly  rules out instantaneous  star formation rates  as those
suggested  by the models.  However, recent  star formation  is obvious
from the  presence of OB  stars as seen  by the ACS/G800L grism  data. 
This dichotomy is reconciled when feedback plays a major role. Delayed
self-regulation  of  star  formation   has  been  known  to  cause  an
overstability that  drives strong  oscillations in the  star formation
rate  (Struck-Marcell  \& Scalo  1987;  Parravano 1996).  Furthermore,
evidence of such intermittent behavior has been presented in our Milky
Way from a detailed study of the chromospheric age distribution of 552
stars in the solar neighborhood  (Rocha-Pinto et al. 2000).  Hence, we
expect  APPLES~1  to  have  a  complex,  intermittent  star  formation
history, which explains the presence of  OB stars and the lack of line
emission. The  chemical enrichment model  presented here gives  just a
``smoothed  view''  of  the  actual formation  process.   The  stellar
continuum -- which we have used  as the main constraint on the stellar
ages  and  metallicities  --  cannot  be  used  to  probe  into  these
oscillations,  whose  characteristic   timescales  are  $\sim  50$~Myr
(Parravano 1996).   Needless to say, the predictions  from this simple
chemical enrichment  model should be  taken as rough estimates  of the
age and metallicity distribution of  the stellar component of the core
of APPLES~1. It should be viewed as a first-order approximation to the
star  formation history  --  the zeroth-order  being  the analysis  of
simple stellar populations.  We want to emphasize that the addition of
a  consistent model  of chemical  enrichment  allows us  to take  into
account  the strong bias  imposed by  the small  $M/L$ of  young stars
whose main  effect is  to reduce luminosity-derived  ages, as  seen in
Figure 12.   More   complicated  models  including  intermittent  star
formation will be more realistic.  However, the effect on the spectral
energy distribution is very hard  to distinguish from a more monotonic
approach  as  the one  shown  above, and  calls  for  a more  detailed
analysis involving {\sl resolved} stellar color-magnitude diagrams.
\par\noindent
One  could still  argue that  young stellar  populations  dominate the
observed SED,  so that  a large fraction  of old (and  low-mass) stars
could go unnoticed by this analysis, implying a large underestimate of
the stellar age. In order to check this point, we ran a grid of models
which combined the best fit  SED with a simple stellar population over
a range  of ages between  8 and 13~Gyr and metallicities  
$-1.7\leq \log(Z/Z_{\odot})\leq -0.2$. A third parameter 
($0\leq f_{\rm O} \leq 1$) controls the  stellar mass fraction in 
this  (old) simple stellar
population. The resulting SEDs  were compared with the observations in
the  same  way  as  described  above,  to find  that  the  best  model
corresponds to $f_{\rm O}=0$. Furthermore,  $f_{\rm O}\geq 0.1$ is 
ruled out at the  $3\sigma$  confidence level. 
Hence  we  can conclude  that  a  significant
contribution to  the {\sl observed spectrum}  of the core  of APPLES~1
from old stars is unlikely.

\section{The environment}

As indicated in Sect.3.1, APPLES~1  is located in the direction of the
Sculptor  Group.  The  canonical members  of this  galaxy group  (i.e. 
NGC~55, NGC~247, NGC~253, NGC~300 and NGC~7793) are characterized by a
radial velocity  between 120 and  230 km~s$^{-1}$ and span  a distance
range between 1.6 and 3.4 Mpc (Puche \& Carignon 1988, Freedman et al.
1992). C\^ot\'e et al. (1997) argued that also NGC~24 and NGC~45 could
belong  to  the Sculptor  Group,  extending it  up  to  a distance  of
$\simeq$ 11 Mpc. The radial velocity measured for NGC~24 and NGC~45 is
V$_{hel}$ =  554 and 468  km~s$^{-1}$ respectively (Puche  \& Carignon
1988). C\^ot\'e  et al. (1997) also  added to the members  list of the
Sculptor  Group 16  dwarf galaxies  (15 gas-rich  systems and  one dS0
galaxy) which cover an interval  in radial velocity between 68 and 702
km~s$^{-1}$.  Their membership  is  discussed on  the  basis of  their
distribution  in  the  (DEC,V$_{LG}$)  plane, where  V$_{LG}$  is  the
velocity  relative to the  Local Group.  In this  same plane  APPLES~1
would  appear to  be a  member of  the Sculptor  Group with  a V$_{LG}
\simeq$ 650 (derived following Yahil  et al. 1977). On the other hand,
if we  set the  Group center  of mass between  NGC~247 and  NGC~253 at
V$_{LG} \simeq$ 270 km~s$^{-1}$ and D $\simeq$ 2.6 Mpc, APPLES~1 would
would lie at $\Delta$$V  \simeq$ 380 km~s$^{-1}$ and $\Delta$$r\simeq$
7 Mpc  from the  center of  the Sculptor Group  and its  crossing time
would  exceed   the  Hubble  time  ($t_{crossing}$   =  2$\Delta$$r  /
\pi\Delta$$V$), whereas  the crossing time  of NGC~247 and  NGC~253 is
about 3 Gyr  (C\^ot\'e et al. 1997).  This result  fully rules out the
possibility that APPLES~1 is a member of the Sculptor Group.

\par

We have extracted  from the NED Database all  the galaxies, except the
Sculptor Group,  with known radial velocity  ($\leq$ 1000 km~s$^{-1}$)
within a  15$^o$ radius from  APPLES~1 (equivalent  to $\simeq$ 2.5 Mpc  at the
fiducial distance  of APPLES  1) to investigate  whether APPLES~1  is a
member of a more distant group  or cluster of galaxies.  Two groups of
galaxies have been  found within the searching radius:  the NBGG 19-06
group  which includes  NGC~7412A, NGC~7424,  NGC~7456 and  NGC~7462 at
V$_r$ = (930 $\pm$ 25) km~s$^{-1}$ (Giuricin et al. 2000), and the LGG
478 group (cf. Garcia 2003, also classified as USGC S289 by Ramella et
al. 2002) which  contains NGC~7713, IC~5332, ESO 347-17  and ESO 348-9
at V$_r$ =  (680 $\pm$ 40) km~s$^{-1}$. Two  field galaxies, ESO 346-7
and ESO 290-28, are also detected  with a radial velocity of about 920
km~s$^{-1}$.  All these galaxies are  plotted as filled dots in Figure
13 where APPLES~1 is represented with the letter ``A''. Larger symbols
indicate  galaxies with  $L \geq  L_*$.  The  left-hand side  panel of
Figure 13 shows  the distribution in RA and DEC  of the above galaxies
around APPLES~1, while the  right-hand side plot shows the distance in
kpc of  the LGG  478 group  from APPLES~1,  under the  assumption that
APPLES~1 is at the group's  distance of $\simeq$ 7.5 Mpc (according to
the LEDA Database, http://leda.univ-lyon1.fr). The distance of APPLES~1 
from the brighter members of the group (IC~5332 and NGC~7713)  is about 700 kpc. 

  To assess  whether APPLES~1 is a plaussible  member of LGC~478 we
  compare this distance to two estimates of its virial radius. For the
  first estimate we simply scale the  size of the Local Group with the
  ratio of  the total B-band  luminosities of LGC~478 ($L  \simeq$ 7.4
  $\times$  10$^9 L_{\odot}$)  and  the Local  Group  ($L \simeq$  4.7
  $\times$ 10$^{10}  L_{\odot}$, computed from the galaxies within
  1 Mpc).  Adopting a  radius of  1 Mpc
  for the  Local Group  this yields  a radius for
  LGC~478 of 160 kpc.  For  our second estimate we use the Conditional
  Luminosity Function (hereafter CLF)  algorithm developed by Yang, Mo
  \& van den  Bosch (2003).  These authors used  the galaxy luminosity
  function and the correlation lenghts as function of luminosity, both
  obtained from the  2 degree Field Galaxy Redshift  Survey, to obtain
  the average mass-to-light ratio $\langle M/L \rangle(M)$ as function
  of halo mass $M$.  Here $L$  is the total luminosity of all galaxies
  that reside in  the same halo. Using CLF models A  to D presented in
  van den  Bosch, Yang  \& Mo (2003),  the total B-band  luminosity of
  LGC~478 suggests an  (average) halo mass of $M_{\rm  vir} \simeq 3.1
  \times 10^{11} h^{-1} M_{\odot}$.  Defining the virial radius as the
  (spherical) radius inside of  which the average overdensity is equal
  to 340 (appropriate for a $\Lambda$CDM cosmology with $\Omega_0=0.3$
  and $\Omega_{\Lambda}=0.7$;  see Bryan \& Norman 1998),  we derive a
  virial radius for LGC~478\footnotemark \/ of 200 kpc (where we have adopted a Hubble
  constant of H$_o$ = 70 km~s$^{-1}$~Mpc$^{-1}$). Both our estimates of
  the virial radius of LGC~478  are in good agreement with each other,
  and more than  three times larger than the  distance to APPLES~1. This
  ratio would turn out even larger when the uncertainty of the distance
  of APPLES~1 is taken into account. This strongly supports the possibility 
  that  APPLES~1 is  a field  dwarf  galaxies, not
  associated with any major group or cluster of galaxies.

\footnotetext{As it can be seen  in the right-hand side plot of Figure
  13, the galaxies  belonging to the LGG 478  group are displaced from
  each  other well  within 200  kpc, the  virial radius  of  the group
  computed  from  their  luminosity  and  the  Conditional  Luminosity
  Function models.  This confirms the  classification of LGG 478  as a
  group  of  galaxies, initially  detected  by  a  Friends of  Friends
  algorithm applied to the coordinates  - velocity plane by Ramella et
  al. (2002) and Garcia (2003).}

\section{Discussion and Conclusions}

Among  the data  collected for  the APPLES  Parallel Program  with the
HST/ACS  Wide Field  Channel, we  have discovered  a  resolved stellar
system whose properties  are consistent with the object  being a dwarf
galaxy. At a  distance of $\simeq$ 9 $\pm$ 2 Mpc (based  on its radial 
velocity and the spectral type of its resolved sources)
APPLES  1 seems to belong to  the field  population  of galaxies and not
associated with any major cluster or group of galaxies.

In the  available ACS  image, APPLES~1  looks quite like  a circularly
symmetric  distribution of  stars, with a  half-light  radius of
  $\simeq$ 1$''.$8 or 78 pc (at the fiducial distance of $\sim$ 9 Mpc)
  and  a total $m_{F775W}$  magnitude of  20.1, which  translates into
  M$_{F775W} \simeq$ -9.6 ($\pm$ 0.5 due to the uncertainty on the
  distance). At face value, the size  of APPLES~1 could
remind  of  the  ultra-compact  dwarf  galaxies  (UCD)  discovered  by
Drinkwater et al. (2003) in the Fornax cluster. Unlike APPLES~1, these
dwarf galaxies are dominated by  old stellar populations (Gregg et al. 
2003) and are believed to be the result of galaxy threshing as modeled
by  Bekki  et  al.   (2001),  whereby nucleated  galaxies  lose  their
envelope  and  are reduced  to  their  only  nucleus by  gravitational
interactions with a more massive galaxy.


  The observed  surface brightness  profile  of APPLES  1 is  well
  fitted by a Sersic profile with a scale length of $\simeq$ 0$''.$35,
  a central surface brightness  of $\simeq$ 21.3 mag~arcsec$^{-2}$ and
  a Sersic index  $n$ = 2.3. Stiavelli et al.   (2001) have shown that
  the Sersic  indices computed for  25 dwarf ellipticals in  the Virgo
  and Fornax clusters range between  0.62 and 4. The authors have also
  pointed out  that the  best fitted galaxies  are characterized  by a
  slope intermediate between  $n$ = 1 and $n$ = 4,  i.e. a range which
  well matches  the Sersic index  of APPLES~1.  The  non-detection of
ionized gas  (i.e.  H$\beta$,  [OIII], H$\alpha$) in  the ACS  and VLT
spectra  together  with  the  spatial  morphology  would  likely
  classify APPLES~1 among the dSph galaxies.  Inquiries of the HIPASS
database show  that no HI  gas has been  detected at the  position and
radial velocity (V$_{hel}$ = 674 km~s$^{-1}$) of APPLES~1. At the
  distance  of APPLES~1,  HIPASS turns  out to  have a  mass detection
  limit of $\simeq$ 10$^8$ M$_o$. So far, accurate HI masses have been
  measured  only for  nearby dwarf  galaxies; for  example,  the dwarf
  ellipticals NGC  185 and NGC 205  have a HI content  of about 10$^5$
  M$_o$ (Young  \& Lo 1997),  while the HI  mass in the  dSph galaxies
  Sculptor  and  Phoenix  ranges   between  10$^3$  and  10$^5$  M$_o$
  (Carignan et al. 1998, St-Germain et al. 1999).


The  comparison  of the  Lick  line indices  as  measured  in the  VLT
spectrum  with the  stellar population  models by  Bruzual  \& Charlot
(2003) gives for the core of APPLES~1 a luminosity-weighted age of 
$1-5$~Gyr.  We have also estimated the age and metallicity of the APPLES~1
core  by fitting  the continuum  slope of  its VLT  spectrum  with the
chemical enrichment models  by Ferreras \& Silk (2000,  2003). In this
way, we have derived a  luminosity-weighted age of $\simeq$ 3 Gyr with
a burst  duration of $\sim$  2 Gyr, a  luminosity-weighted metallicity
$\log Z/Z_\odot=-0.2$~dex with a  metallicity spread of $\simeq 0.2$~dex. 
The best  fitting model gives  for the  stellar population  dominating the
optical  a mass-to-ligh  ratio M/L$_V$  = 1.4  and a  stellar  mass of
$\simeq 8\times 10^{6}$ M$_{\odot}$.  The infall of  gas, which
feeds the ongoing burst, has apparently started at a redshift $z_F\sim 3$ 
and has a  timescale of about 5~Gyr.  The efficiency at which  the core of
APPLES~1  forms stars  is about 0.7~Gyr$^{-1}$, typical  of late-type
galaxies.
The presence of OB stars and an integrated spectra which matches an
F type star argues against a simple stellar population, whose best fit
requires a very young age ($\lesssim 1$~Gyr) and supersolar metallicities,
which are in conflict with the weakness of the \ion{Ca}{2} $H$ 
and $K$ lines, and with the relatively weak Mgb feature. Hence, 
the spectra strongly suggests an extended period of star formation
over several Gyr.

  As  shown in Mateo (1998)  and Grebel (1997),  the star formation
  history of dSphs  in the Local Group is  quite complicated. Five out
  of  23  dSphs listed  in  Mateo (1998)  present  one  burst of  star
  formation, now observed in the  form of RGB stars.  The remaining 18
  dSphs  are characterized  by multiple  bursts, seven  of  which show
  main-sequence stars with  age = 0 Gyr. Bursts  like the one observed
  in APPLES~1  (i.e. dominated by a 3 Gyr  old stellar population) are
  known to  occur in a  number of dSph  and dIrr/dSph galaxies  in the
  Local Group: NGC~205, Fornax,  Carina, Sextans, DDO~210 and Pegasus. 
  In particular,  the VLT spectrum  of the core  of APPLES~1  is quite
  similar to  that of the  core of NGC~205  (Bica et al. 1990)  with a
  stellar  population  comparably  metal-rich  and  young.  Therefore,
  APPLES~1  would be  consistent with the  dSph galaxies in  the Local
  Group. But there are four important differences with the Local Group
  dSphs:
\par\noindent
{\it i)} APPLES~1 is apparently an isolated system, 
\par\noindent
{\it ii)} two nearby dSphs, Tucana and Cetus, very likely unbound
to the Local Group, do not show any recent burst of star formation and their stellar 
populations date back to $\sim$ 10 Gyr ago (Mateo 1998, Grebel 1997). 
\par\noindent
{\it iii)} NGC~205 probably was a disc galaxy and its interaction with M31 has removed
its gas with the net effect of ``shrinking'' NGC~205 to its bulge.
\par\noindent
{\it iv)} no old stellar component has been detected in the data currently
available for APPLES~1.
\par\noindent
Tidal interactions, ram-pressure, collisions, merger events and ionization 
from the UV background are thought to regulate the depletion of gas (primordial and 
ejected by stars during their evolution) and the star formation in satellite dSphs (cf. 
Grebel et al. 2003, Dong et al. 2003). Therefore, 
Grebel et al. (2003) suggest that Tucana and Cetus could be runaways 
or have been stripped of their gas by ram-pressure along their orbit if the intergalactic 
medium of the Local Group were clumpy. Alternatively, these dwarf galaxies could have been more
massive in the past and had been heated and truncated by the tidal interactions with
the massive neighbors at high redshift, as suggested by Kravtsov et al. (2004b). 
Certainly, a better knowledge of their orbital parameters would 
greatly help in understanding their origin. If these ruled out the occurrence of any major
interactions with the Milky Way, then Tucana and Cetus would have evolved as predicted
by Dekel \& Silk (1986) with supernova explosions removing their gas. 
The key point, here, is to determine the full 
star formation history experienced by APPLES~1. For example,
the best fitting model of chemical enrichment (as well as the Lick line indices)
indicates that the VLT spectrum of the core of APPLES~1 is dominated by young and
intermediate-age stars with a negligible contribution from old stars to the
optical light. This does not rule out that old stars, if present, could contribute
significantly to the total mass of APPLES~1. 
Is there a $\sim$ 10 Grys old population in APPLES~1 as typically observed in dwarf 
galaxies in the Local Group, or is APPLES~1 experiencing the very first burst of
star formation similarly to GR~8, a dIrr galaxy in the Local Group with a stellar
population younger than 1 Gyr (Mateo 1998)? In this second hypothesis, APPLES~1 would
resemble a dwarf galaxy in the predictions of Dekel \& Silk (1986) which formed out
of the intergalactic medium (IGM) at $z$ = 3 and has lost its
gas through stellar winds after a single burst of star formation.
The metallicity derived for APPLES~1, though, would require a significantly enriched IGM, 
whereas the intergalactic metallicity measured at $z \simeq$ 3
is about -1 dex (Pettini 2004). 
Significantly deeper and multi-wavelength 
imaging with HST is necessary to detect the older and fainter stars to re-construct the full
star-formation history of APPLES~1. The same observations will also provide a more
accurate measure of the distance of this galaxy. If these observations confirm the
distance estimated for APPLES~1 in this paper and reveal the presence/absence of older
stellar components, they will set the challenge of explaining what regulates
star formation in a isolated dSph galaxy.

\acknowledgments
We would like to thank N. Huelamo and O. Marco for carrying out the 
VLT/FORS2 observations in servicing
mode and M. Petr-Gotzens as ESO contact scientist.
AP would like to thank F. van den Bosch, J.S. Gallagher, E.K. Grebel
and A. Tarchi for stimulating discussions. We acknowledge the helpful
suggestions and comments from an anonymous referee which certainly
improved the paper. Support for the APPLES project at STScI has
been provided under NASA HST grant GO-09482.01-A. 
The Hubble Space Telescope is a project of international
cooperation between NASA and the European Space Agency (ESA).
The Space Telescope Science Institute is operated by the Association
of Universities for Research in Astronomy, Inc., under NASA Contract
NAS5-26555. This research has made use of the NASA/IPAC Extragalactic Database 
(NED) which is operated by the Jet Propulsion Laboratory, California Institute 
of Technology, under contract with the National Aeronautics and Space Administration.

\clearpage
\begin{table}
\caption{Best composite model for APPLES~1 ($3\sigma$ confidence level)}
\begin{center}
\begin{tabular}{cc}
Chemical enrichment parameters & Physical parameters\\
\hline
$\beta < 0.20$ & $\langle t\rangle = 3.27^{+0.50}_{-0.60}$~Gyr\\
$\nu = 0.68^{+2.44}_{-0.36}$~Gyr$^{-1}$ & $\sigma_t = 2.18^{+0.23}_{-0.40}$~Gyr\\ 
$z_F = 2.7^{+0.3}_{-1.4}$ & $\langle [Z/H]\rangle = -0.18^{+0.28}_{-0.23}$\\
$\tau_f > 4.9$~Gyr & $\sigma_{[Z/H]}=0.23^{+0.04}_{-0.09}$\\
$\chi^2_r = 1.14$ & $M/L_V=1.39^a$\\
deg.~of~freedom$=235$ & $\log M_\star/M_\odot \sim 6.9^b$\\
\hline\hline
$^a$ Assuming a Salpeter IMF.
$^b$ For $M_V=-9$.
\end{tabular}
\end{center}
\end{table}

\clearpage

\begin{figure}
\epsscale{1.0}
\plotone{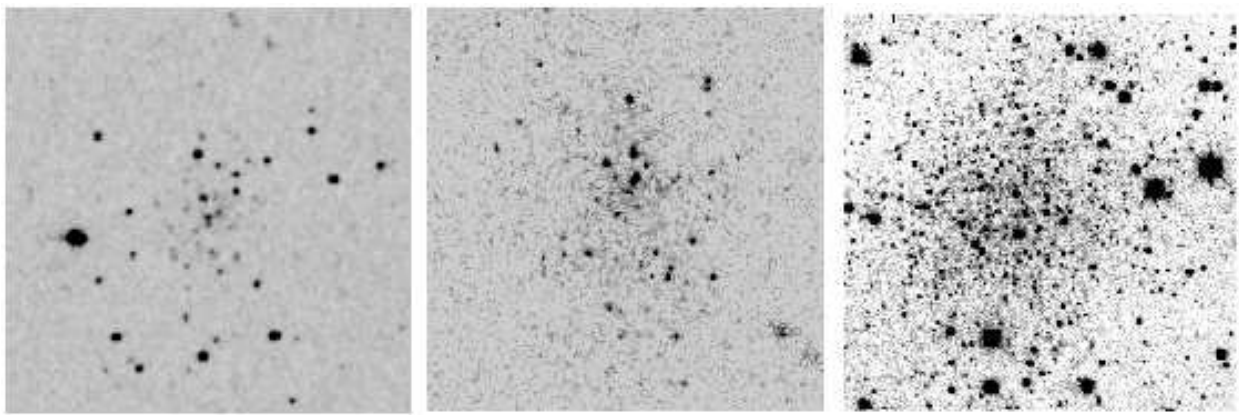}
\caption{The F775W image of APPLES~1 (middle panel, 9$'' \times$ 9$''$, i.e. 
$\sim$ 440 $\times$ 440 pc$^2$) compared with the DSS image of the globular cluster
Pal 13 in the Galactic halo (left-hand side panel, 5$' \times$ 5$'$, i.e. $\sim$ 130 $\times$ 130
pc$^2$) and a V-band image of the dwarf galaxy And V (right-hand
side panel, 3$' \times$ 3$'$ equivalent to 640 $\times$ 640 pc$^2$) by Armandroff et al. (1998).}
\end{figure}

\begin{figure}
\epsscale{0.7}
\plotone{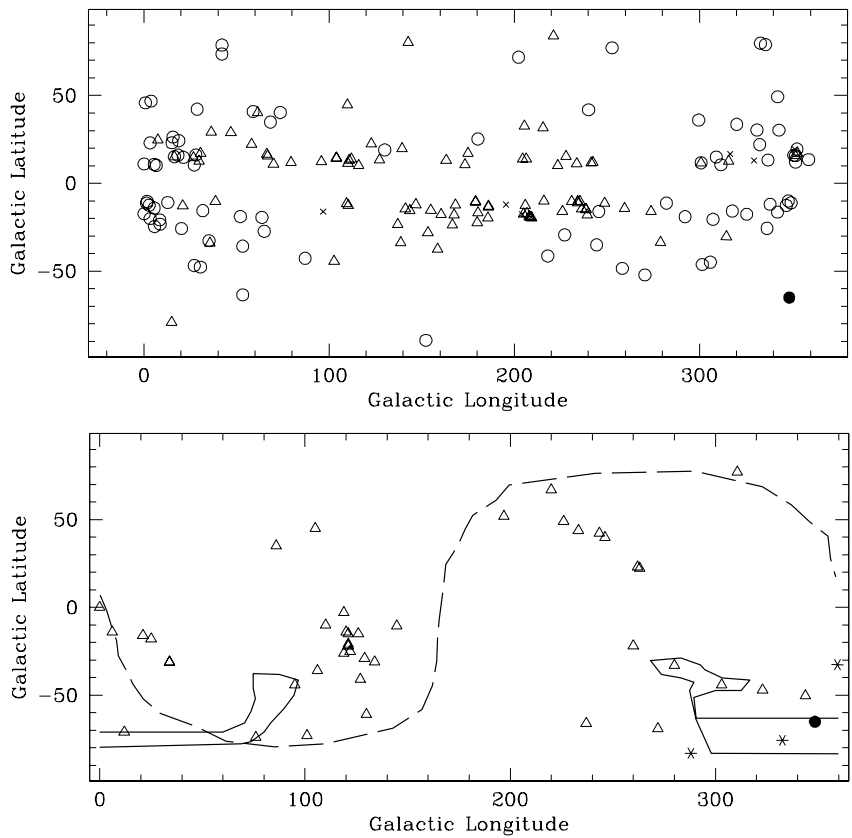}
\caption{{\it Top:} the spatial distribution, in galactic {\it (l,b)} 
coordinates of Galactic OB associations (crosses), open clusters (open triangles)
and globular clusters (open circles). APPLES~1 is represented with a filled circle.
{\it Bottom:} the spatial distribution of the members of the Local Group (open triangles).
The galaxies of the Sculptor
Group appear as asterisks and APPLES~1 is indicated with a filled
circle. The dashed and solid lines trace the Sagittarius (Martinez-Delgado et
al. 2003) and Magellanic (Kunkel 1979) streams, respectively.}
\end{figure}

\begin{figure}
\epsscale{0.5}
\plotone{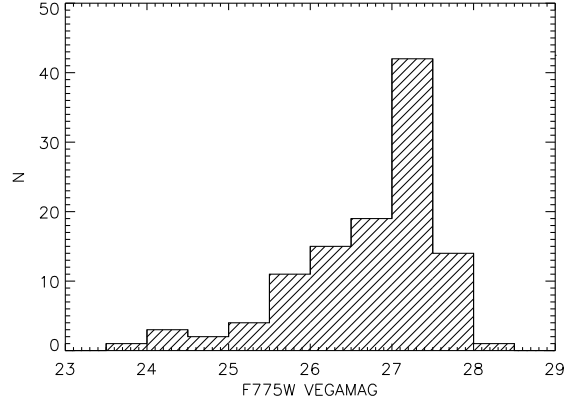}
\caption{The observed luminosity function of the sources detected in
APPLES~1.}
\end{figure}

\begin{figure}
\epsscale{0.5}
\plotone{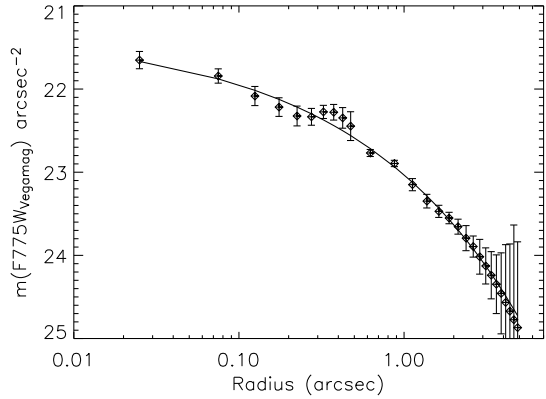}
\caption{The observed surface brightness profile of APPLES~1. The solid line
represents the best fitting Sersic profile, obtained for a scale length
$r_s$ = 0$''.$35 $\pm$ 0$''.$14, a central surface brightness $\Sigma_o$ 
= 21.33 $\pm$ 0.18 mag~arcsec$^{-2}$ and a Sersic index $n$ = 2.30 $\pm$ 0.29.}
\end{figure}

\begin{figure}
\epsscale{0.7}
\plotone{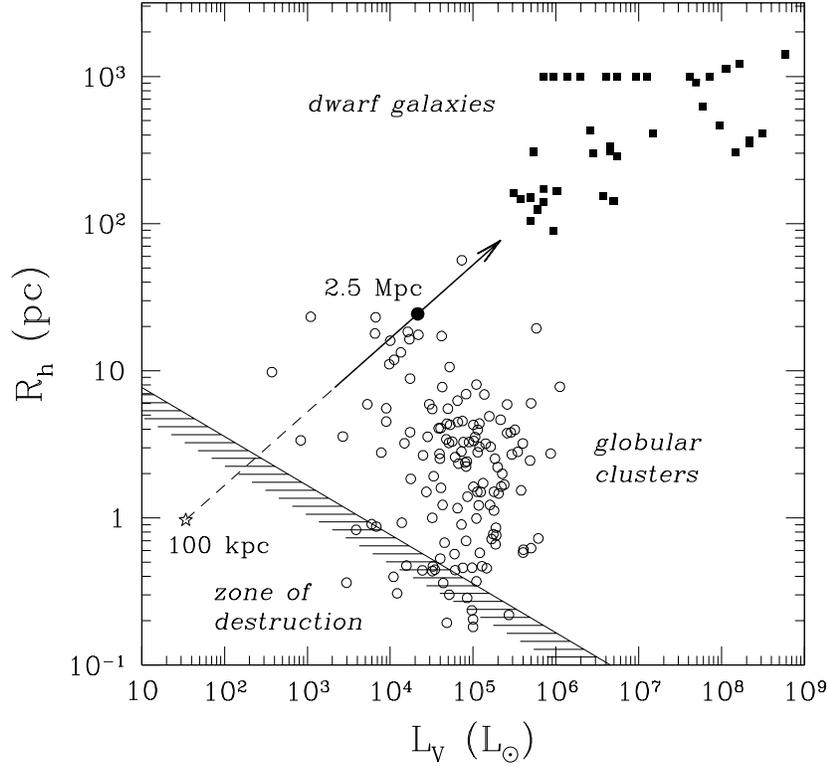}
\caption{Location of the stellar system on the
luminosity in V band - size plane (filled circle) along with the Galactic globular
(open circles) and dwarf galaxies of the Local Group (filled
squares).  The location is set at an estimated distance of 2.5 Mpc;
the arrow shows the effect of changing the distance by a factor of 3
and points in the direction of increasing distance.  Dashed line
points towards the hypothetical distance of 100 kpc (marked by
asterisk), the largest distance for a Galactic object.  In the ``zone
of destruction'' an isolated cluster would be disrupted in the Hubble
time due to two-body relaxation.}
\end{figure}

\begin{figure}
\epsscale{0.5}
\plotone{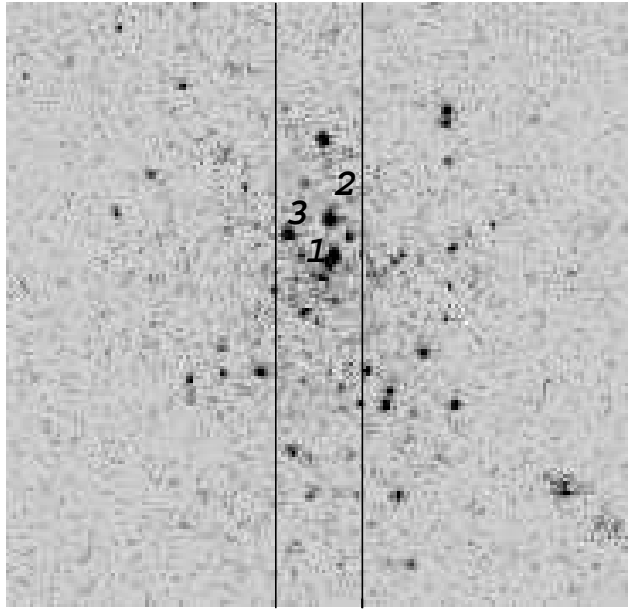}
\caption{The position of Sources $\#$ 1, 2 and 3 in the ACS F775W image of 
APPLES~1. The field of view is 9$'' \times$ 9$''$. The slit, 1$''$.6 wide,
used for the VLT/FORS2 observations is also drawn.}
\end{figure}

\begin{figure}
\plotone{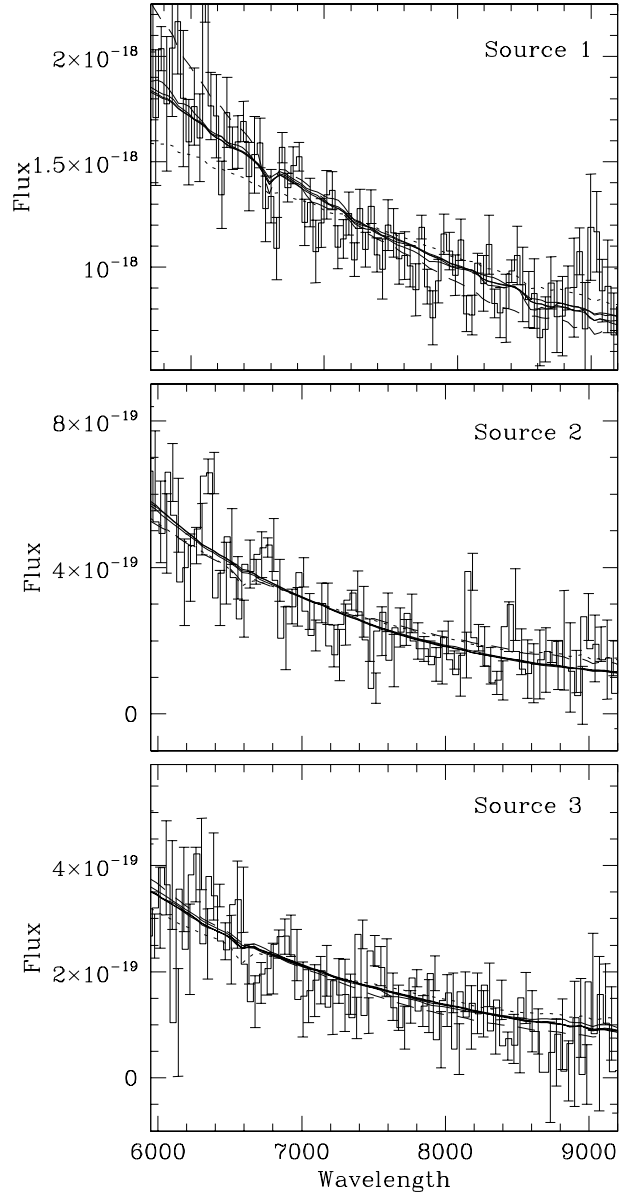}
\caption{The G800L grism spectra of the brightest sources numbered in 
Figure 6. Solid lines represent the best-fit templates which vary from 
the spectral type F5 - F8 for Source $\#$ 1, to O5 - O9 for Source $\#$ 2
and B3 - B6 for Source $\#$ 3. Fluxes are in units of 
erg~s$^{-1}$~cm$^{-2}$~\AA$^{-1}$ and wavelengths are in \AA.} 
\end{figure}

\begin{figure}
\epsscale{1.0}
\plotone{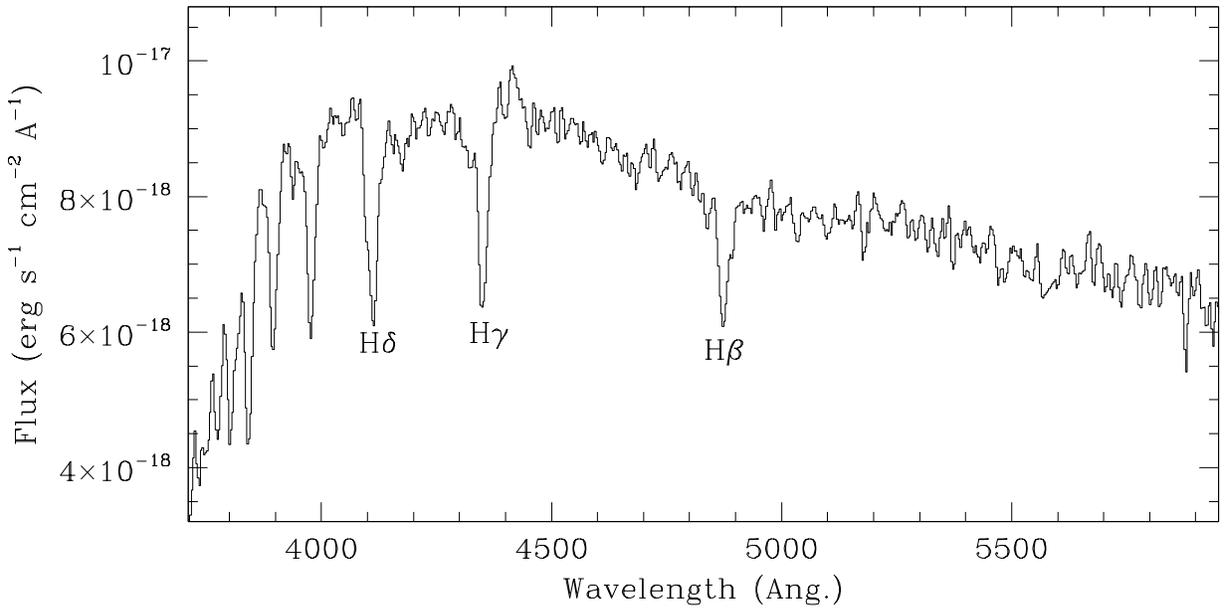}
\caption{A smoothed version of the VLT/FORS2 spectrum of APPLES~1. The average flux uncertainty is 7$\%$.}
\end{figure}

\begin{figure}
\epsscale{1.0}
\plotone{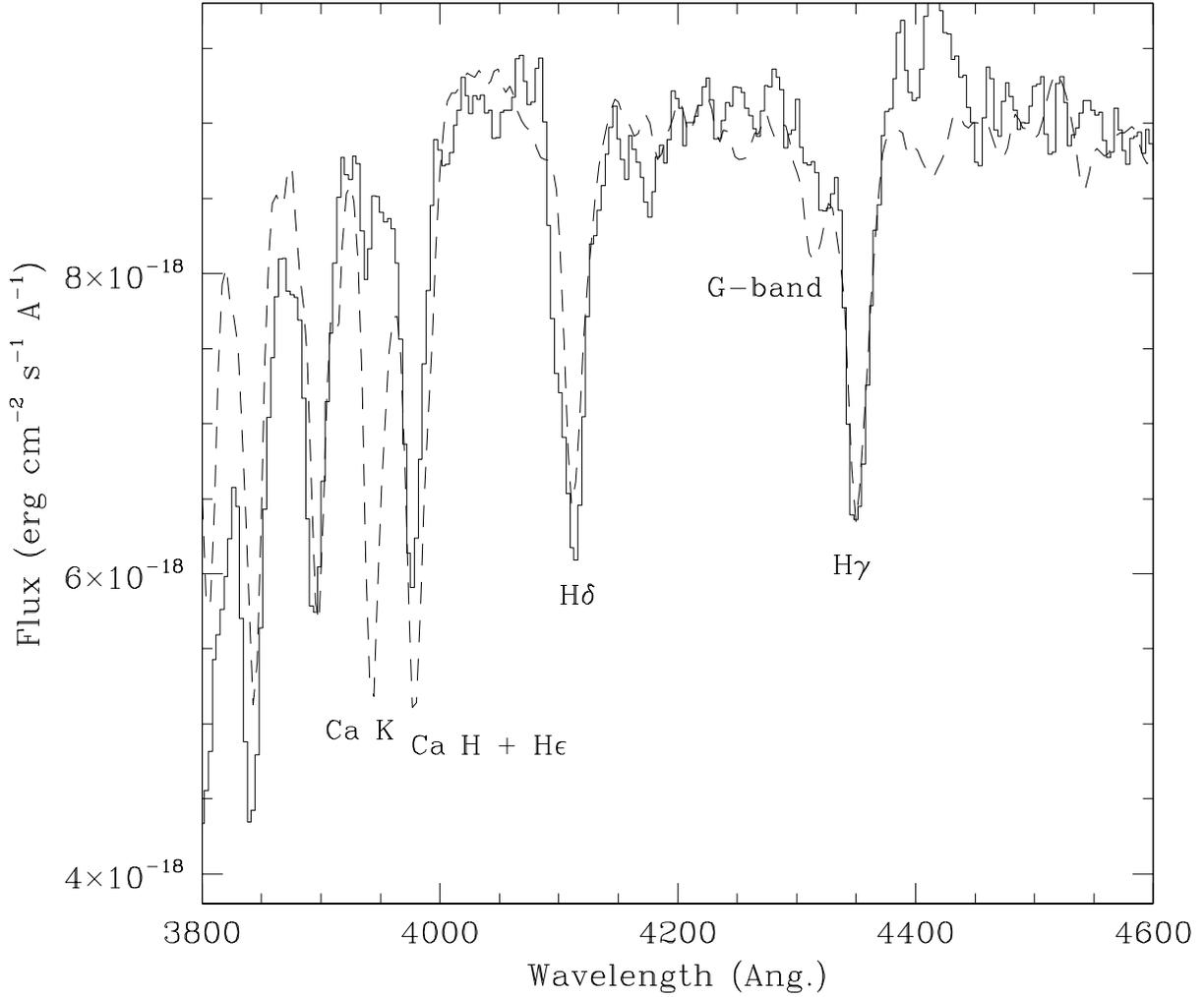}
\caption{The VLT/FORS2 spectrum of APPLES~1 (solid line) is fitted with a combination
of Pickles' F2V and F5V spectral types. Note the discrepancy in the Ca K line between
the observed and the template spectra.}
\end{figure}

\begin{figure}
\epsscale{0.8}
\plotone{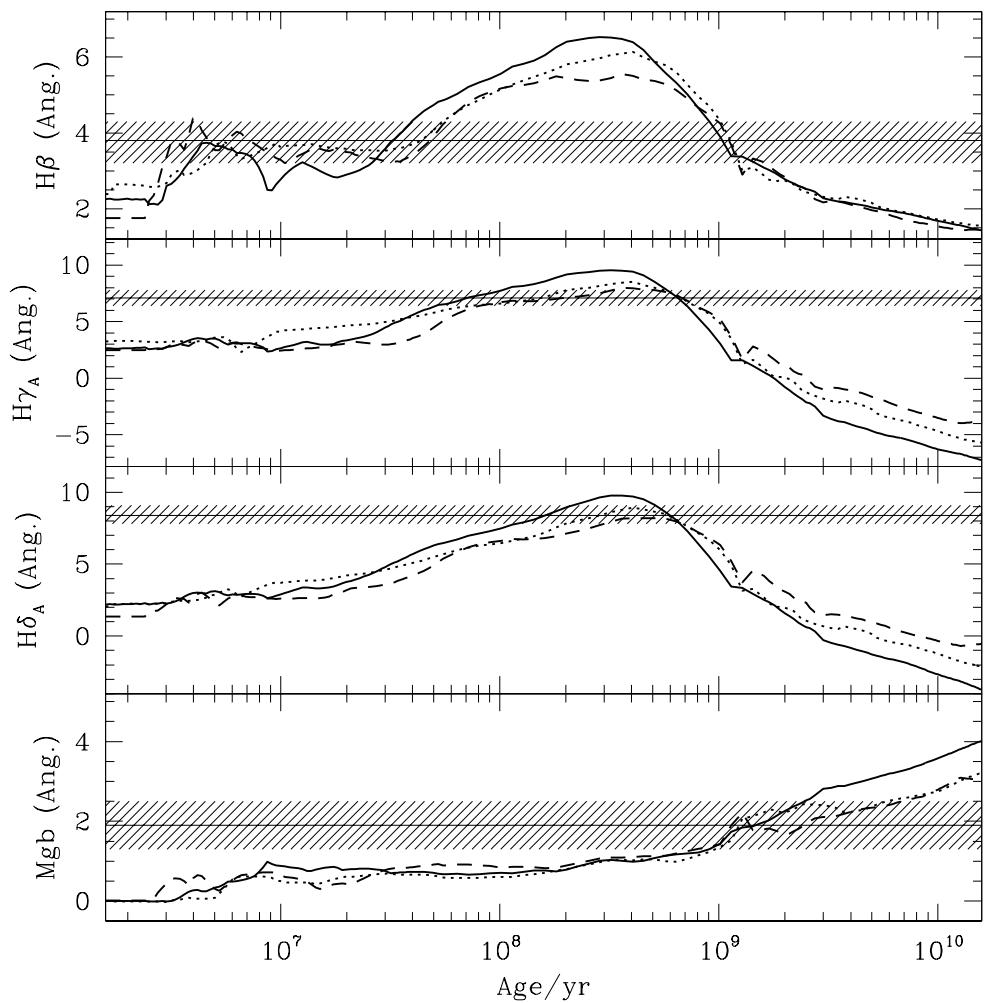}
\caption{The Lick H$\beta$, H$\gamma_A$, H$\delta_A$ and Mgb indices measured for APPLES~1
and plotted against the Bruzual \& Charlot (2003) models. The solid, dotted
and dashed lines indicate models with solar, LMC and half LMC metallicity respectively.
The horizontal solid line corresponds to the measured Lick index and the shaded areas
visualize the 1 $\sigma$ confidence levels.}
\end{figure}

\begin{figure}
\epsscale{0.7}
\plotone{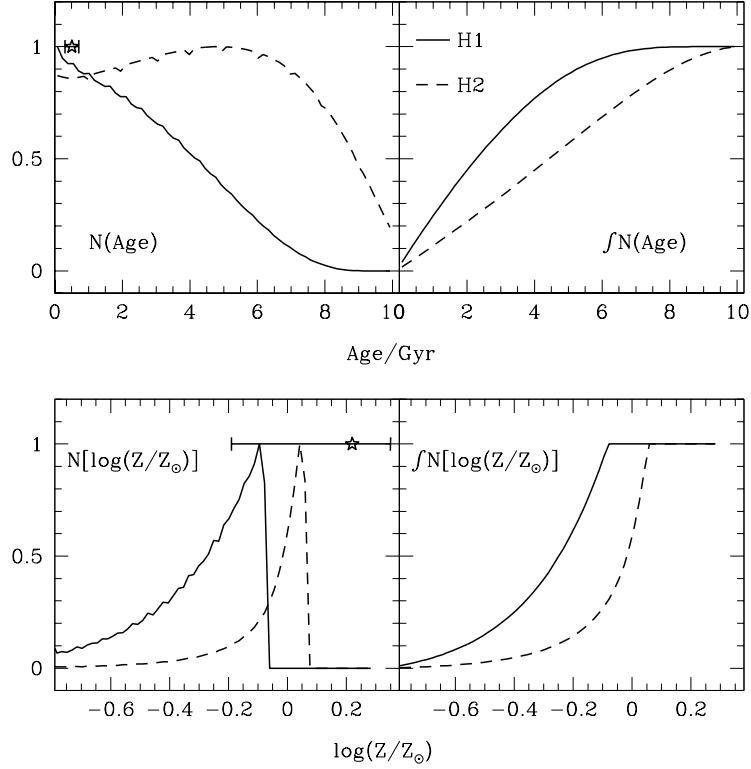}
\caption{ Age and metallicity distribution for a range
of star formation histories that give the best fit to
the VLT/FORS2 spectrum of APPLES~1. The left panels give
the actual distributions of age ({\sl top}) and metallicity
({\sl bottom}), whereas the panels on the right give the
cumulative distributions. Histories H1 and H2 represent a
wide range of parameters (see text for details). These two histories 
span the $3\sigma$ confidence level. The stars at the top of the panels 
on the left give the best fit and $3\sigma$ confidence levels for a
similar comparison of the spectrum of APPLES~1 with a grid of simple stellar
populations (single age and metallicity) from the Bruzual \& Charlot (2003)
models.
}
\end{figure}

\begin{figure}
\epsscale{1.0}
\plotone{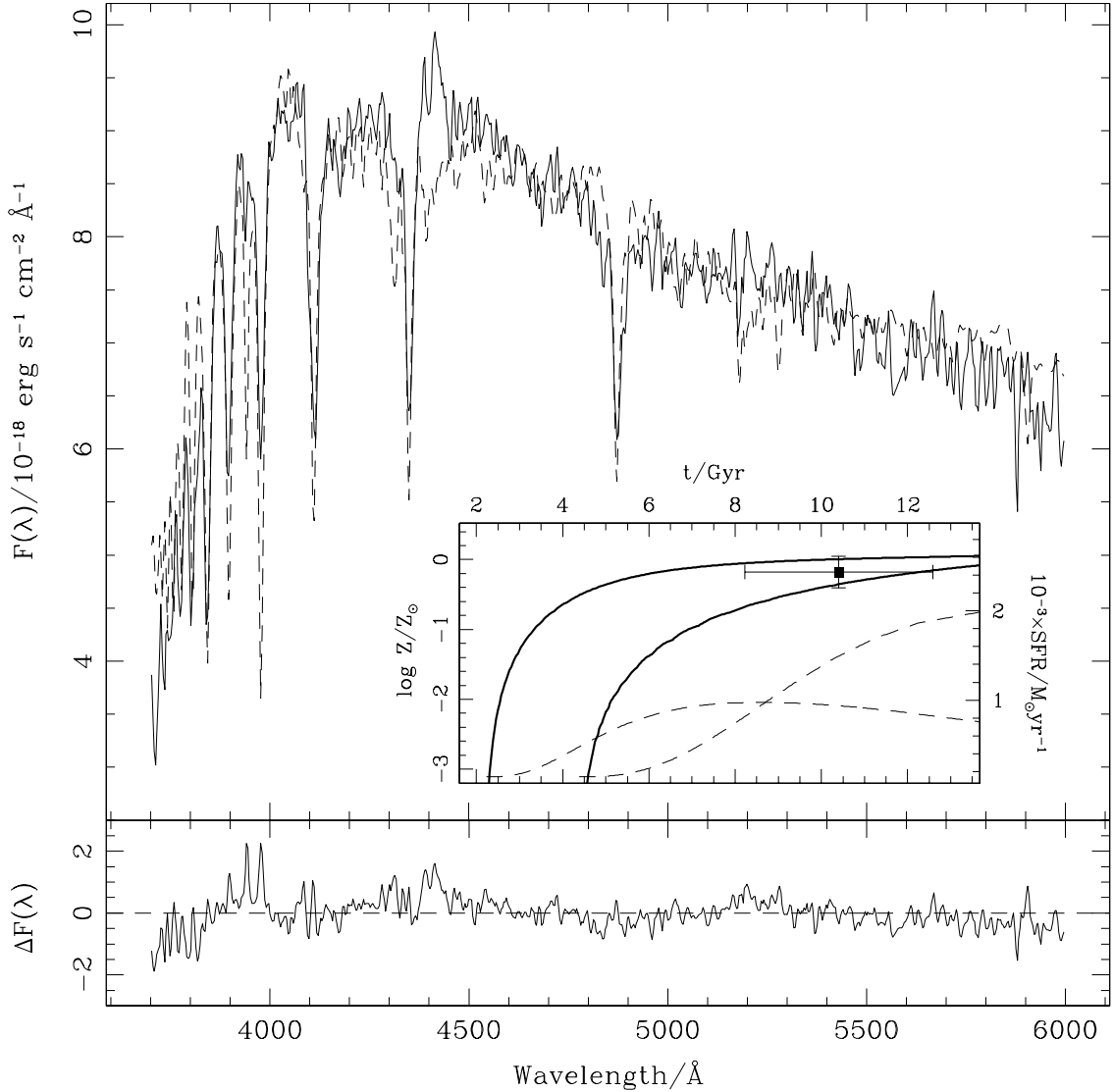}
\caption{Best fit of the composite model (dashed line) compared with
the VLT/FORS2 spectrum of APPLES~1. The bottom panel gives the
residuals as a function of the uncertainty for each spectral element
(in the same units, namely $10^{-18}$~erg~s$^{-1}$~cm$^{-2}$~\AA$^{-1}$).
The inset gives the star formation history corresponding to the
best fit. We show the star formation rate as a dashed line and the
instantaneous metallicity of the ISM as a solid line. The star formation
rates are normalized to an absolute luminosity of $M_V=-9$ at z=0.
Two star formation histories are shown, 
encompassing the allowed range of parameters
within a $3\sigma$ confidence level. The point gives the
average age and metallicity of the best fit and the error bars
show the RMS.}
\end{figure}

\begin{figure}
\epsscale{1.0}
\plotone{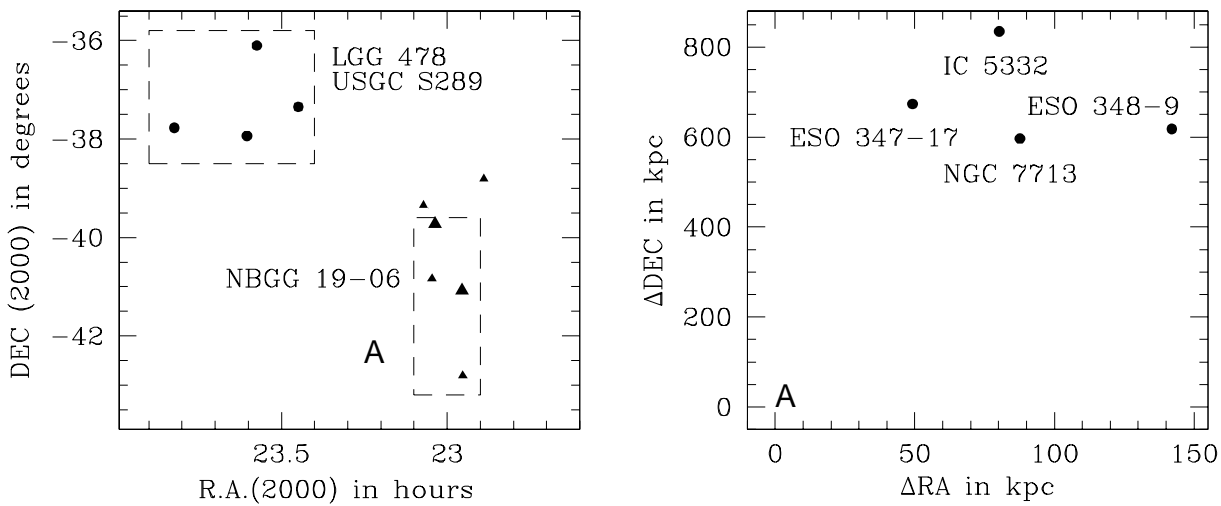}
\caption{The galaxies with a radial velocity smaller than 1000 km~s$^{-1}$
in a 15$^o$ radius from APPLES~1 (indicated with the letter ``A'') are
plotted on the plane of the sky in the left-hand side panel. The group
of galaxies closest to APPLES~1 in terms of radial velocity (LGG 478, USGC S289)
is plotted in the right-hand side panel in units of linear distance from
APPLES~1, under the assumption that LGC 478 and APPLES~1 are at the same 
distance of $\sim$ 7.5 Mpc.}
\end{figure}

\end{document}